\documentclass[%
preprint,
amsmath,amssymb,
aps,
]{revtex4-1}

\usepackage{graphicx} 
\graphicspath{ {Figures/} }
\usepackage{epstopdf}
\usepackage[caption=false]{subfig}
\usepackage{amsmath}
\usepackage{amsfonts}
\usepackage{amssymb}
\usepackage{bm}
\usepackage{hyperref}
\usepackage{amsthm}
\usepackage{color}

\usepackage{mathrsfs}
\usepackage[vlined,ruled]{algorithm2e}
\usepackage{url}
\usepackage{color}
\usepackage{algorithmic}
\usepackage{bbm}
\usepackage{booktabs}
\usepackage{array}
\usepackage[table]{xcolor}
\usepackage{yfonts}
\usepackage{lineno}

\newcommand\oprocendsymbol{\hbox{$\square$}}
\newcommand\oprocend{\relax\ifmmode\else\unskip\hfill\fi\oprocendsymbol}

\begin{document}
\bibliographystyle{naturemag}
\title{A Neural Programming Language for the Reservoir Computer}
\author{Jason Z. Kim}
\affiliation{Department of Bioengineering, University of Pennsylvania, Philadelphia, PA, 19104}
\author{Dani S. Bassett}
\affiliation{Departments of Bioengineering, Physics \& Astronomy, Electrical \& Systems Engineering, Neurology, and Psychiatry, University of Pennsylvania, Philadelphia, PA, 19104}
\affiliation{Santa Fe Institute, Santa Fe, NM 87501}
\affiliation{To whom correspondence should be addressed: dsb@seas.upenn.edu}
\date{\today}

\begin{abstract}
	From logical reasoning to mental simulation, biological and artificial neural systems possess an incredible capacity for computation. Such neural computers offer a fundamentally novel computing paradigm by representing data continuously and processing information in a natively parallel and distributed manner. To harness this computation, prior work has developed extensive training techniques to understand existing neural networks. However, the lack of a concrete and low-level programming language for neural networks precludes us from taking full advantage of a neural computing framework. Here, we provide such a programming language using reservoir computing---a simple recurrent neural network---and close the gap between how we conceptualize and implement neural computers and silicon computers. By decomposing the reservoir's internal representation and dynamics into a symbolic basis of its inputs, we define a low-level neural machine code that we use to program the reservoir to solve complex equations and store chaotic dynamical systems as random access memory (dRAM). Using this representation, we provide a fully distributed neural implementation of software virtualization and logical circuits, and even program a playable game of pong inside of a reservoir computer. Taken together, we define a concrete, practical, and fully generalizable implementation of neural computation.
\end{abstract}
\maketitle

\section{Introduction}
Neural systems possess an incredible capacity for computation. From biological brains that learn to manipulate numeric symbols and run mental simulations \cite{nieder2009representation,salmelin1994dynamics,hegarty2004mechanical} to artificial neural networks that are trained to master complex strategy games \cite{silver2016mastering,silver2017mastering}, neural networks are outstanding computers. What makes these neural computers so compelling is that they are exceptional in different ways than modern-day silicon computers: the latter relies on binary representations, rapid sequential processing \cite{patterson2016computer}, and segregated memory and CPU \cite{von1993first}, while the former utilizes continuum representations \cite{singh2017consensus,gollisch2008rapid}, parallel and distributed processing \cite{sigman2008brain,nassi2009parallel}, and distributed memory \cite{rissman2012distributed}. To harness these distinct computational abilities, prior work has studied a vast array of different network architectures \cite{cho2014properties,towlson2013rich}, learning algorithms \cite{werbos1990backpropagation,caporale2008spike}, and information-theoretic frameworks \cite{tishby2000information,olshausen1996emergence,kline2021gaussian} in both biological and artificial neural networks. Despite these significant advances, the relationship between neural computers and modern-day silicon computers remains an analogy due to our lack of a concrete and low-level programming language, thereby limiting our access to neural computation.

To bring this analogy to reality, we seek a neural network with a simple set of governing equations that demonstrates many computer-like capabilities \cite{lukovsevivcius2012reservoir}. One such network is a reservoir computer (RC), which is a recurrent neural network (RNN) that receives inputs, evolves a set of internal states forward in time, and outputs a weighted sum of its states \cite{jaeger2001echo,sussillo2009generating}. True to its namesake, reservoir computers can be trained to perform fundamental computing functions such as memory storage \cite{lu2018attractor,kocarev1996generalized} and manipulation \cite{smith2022learning,kim2021teaching}, prediction of chaotic systems \cite{sussillo2009generating}, and model-free control \cite{canaday2021model}. Owing to the simplicity of the governing equations, the theoretical mechanism of training is understood, and recent advances have dramatically shortened training requirements by using a more efficient and expanded set of input training data \cite{gauthier2021next}. But can we skip the training altogether and program RCs without running a single simulation just as we do for silicon computers? Combined with the substantial advances in experimental RC platforms \cite{tanaka2019recent}, it is now timely to formalize a programming language to develop neural software atop RC hardware.

Here, we provide such a programming language by formally constructing a symbolic representational basis of the RC neurons. First we write the state of the RC neurons as a symbolic sum of polynomials in the input variables and their time derivatives, and use this sum to program symbolic outputs. We then show that performing feedback by using the outputs to drive the inputs produces an equation of equivalence between the outputs and the inputs, such that the time evolution of the RC naturally solves the equation in a natively continuous and parallel manner. We use this feedback to solve the least-squares regression problem and the Lyapunov equation for the controllability Gramian using RCs. Then, we expand our programming language to also include the dynamics of the output by accounting for the time derivative of the RC. 

Using this expansion, we turn the analogy between neural computation and silicon computation into a concrete reality by programming fundamental constructs from computer science into RCs. First, we extend the idea of static memory in silicon computers to program chaotic dynamical systems as random access memories (dRAM). Second, because RCs can store dynamical systems as memories, and the RC itself is a dynamical system, we demonstrate that a host RC can \textit{virtualize} the time-evolution of a guest RC, precisely as a host silicon computer can create a virtual machine of a guest computer. Third, we provide a concrete implementation of a fully neural logical calculus by programming RCs to evolve as the logic gates \textsc{and, nand, or, nor, xor}, and \textsc{xnor}, and construct neural implementations of logic circuits such as a binary adder, flip-flop latch, and multivibrator circuit. Finally, we define a simple scheme for software and game development on RC architectures by programming an RC to simulate a variant of the game ``pong.'' Through this language, we define a concrete, practical, and fully generalizable implementation of neural computation.

\section{Symbolic Decomposition of Neural Representation}
\begin{figure}[h!]
	\includegraphics[width=\textwidth]{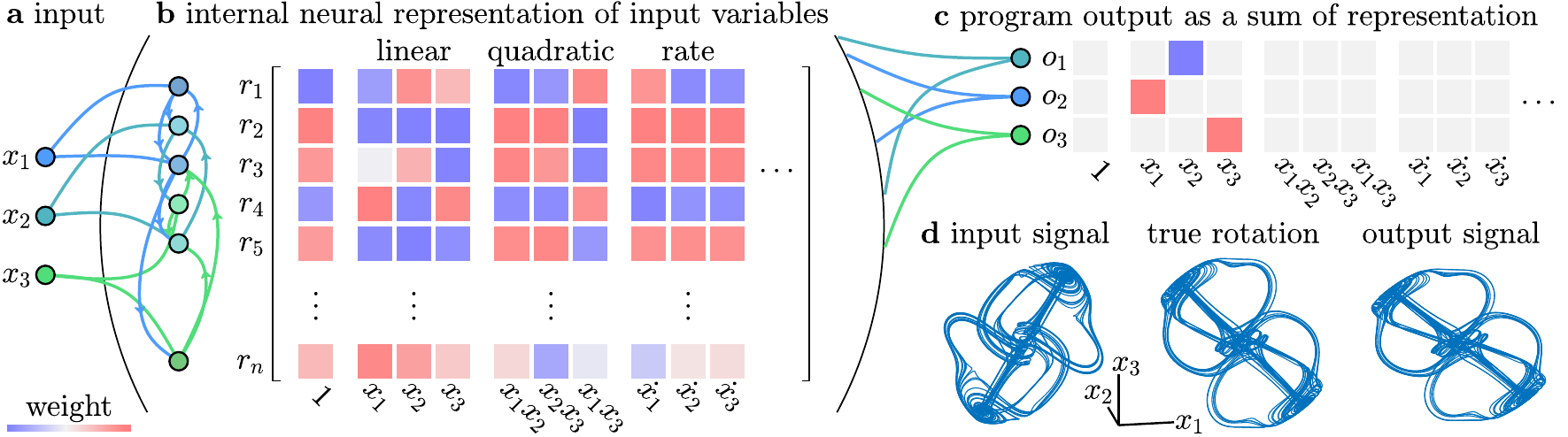}
	\caption{\label{fig:decomposition}\textbf{Unfurling neural states as a weighted sum of input variables.} (\textbf{a}) Inputs to our RNN, which do not represent specific numerical values, but rather symbolic variables. (\textbf{b}) We expand the activity of the RNN neurons as a weighted sum of polynomials in the input variables and their time derivatives. (\textbf{c}) We can then program an output matrix $W$ that maps the RNN's symbolic representation of its inputs to any analytical function of the inputs, such as a rotation. (\textbf{d}) When we drive the programmed RNN with a complex input such as the chaotic Thomas attractor, the output indistinguishably rotates the input (typical relative error is less than 1\%).}
\end{figure}

To construct such a language, we begin with a simple equation for our RNN comprising $n$ neurons and $m$ inputs:
\begin{equation}
\label{eq:RNN}
\frac{1}{\gamma} \dot{\bm{r}}(t) = -\bm{r}(t) + \bm{g}(A\bm{r}(t) + B\bm{x}(t) + \bm{d}).
\end{equation}
Here, $\bm{r} \in \mathbb{R}^n$ are the neuron states, $\bm{x} \in \mathbb{R}^m$ are the input states, $A \in \mathbb{R}^{n\times n}$ are the connections between neurons, $B \in \mathbb{R}^{n \times m}$ are the connections from inputs to neurons, $\bm{d} \in \mathbb{R}^n$ are the bias terms, $\bm{g}$ is a sigmoidal function that acts element-wise on its inputs (thereby mapping vectors to vectors), and $\gamma$ is the time constant. In short, Eq.~\ref{eq:RNN} defines the time evolution of neural states as a function of the current neural and input states. For conciseness in what follows, we will omit the notation $(t)$ that explicitly denotes time when obvious.

To define our programming language, we would like to write $\bm{r}$ as a simple function of $\bm{x}$, which is difficult to do for nonlinear systems. Hence, we linearize the system about the neural state $\bm{r}^*$, and partially linearize the system about the input state $\bm{x}^*$ to yield
\begin{equation}
\label{eq:RNN_linear}
\frac{1}{\gamma}\dot{\bm{r}}(t) = A^*\bm{r}(t) + \bm{u}(\bm{x}(t)),
\end{equation}
where $A^*$ is the effective dynamical response as $\bm{r}(t)$ deviates slightly away from $\bm{r}^*$, and $\bm{u}(\bm{x}(t))$ is the effective input into the system (see Supplemental Section I for the derivation). 

Next, because Eq.~\ref{eq:RNN_linear} is linear, we can write the neural states $\bm{r}(t)$ as a convolution of the input with the impulse response. This convolution involving integrals can be expressed algebraically by taking the series expansion of $\bm{u}(\bm{x}(t))$ with respect to time, to yield
\begin{equation}
\label{eq:RNN_representation}
\begin{split}
\bm{r}(t)
&= \bm{h}(\bm{x}(t),\dot{\bm{x}}(t),\ddot{\bm{x}}(t),\dotsm),
\end{split}
\end{equation}
where $\bm{h}$ is an analytic function that we derive in Supplemental Section I. Finally, we perform a Taylor series expansion of $\bm{h}$ with respect to $\bm{x},\dot{\bm{x}},\dotsm$ to decompose the reservoir state $\bm{r}$ as a weighted sum of a polynomial basis of input variables (Fig.~\ref{fig:decomposition}a,b). It is precisely the coefficients of this expansion---shown in Fig.~\ref{fig:decomposition}b---that form the start of our recurrent neural \textit{programming language}. For example, we can program the RNN to output a rotation of its inputs by solving for an output matrix $W \in \mathbb{R}^{k \times 3}$ that rotates any input about the $x_3$ axis (Fig.~\ref{fig:decomposition}c). When we evolve the RNN with the chaotic Thomas attractor given by
\begin{equation*}
\begin{split}
\dot{x}_1(t) &= -3.7x_1(t) + 5\sin(4x_2(t))\\
\dot{x}_2(t) &= -3.7x_2(t) + 5\sin(4x_3(t))\\
\dot{x}_3(t) &= -3.7x_3(t) + 5\sin(4x_1(t)),
\end{split}
\end{equation*}
we find that the output is indistinguishable from a true rotation of the input (Fig.~\ref{fig:decomposition}d). See Supplemental Section I for a detailed derivation, Supplemental Sections II and III for an explicit quantification of the goodness of the approximation as a function of the RC parameters, and Supplemental Section IV for the parameters used for all examples.

\section{Natively Distributed and Symbolic Computation}
\begin{figure}[h!]
	\includegraphics[width=\textwidth]{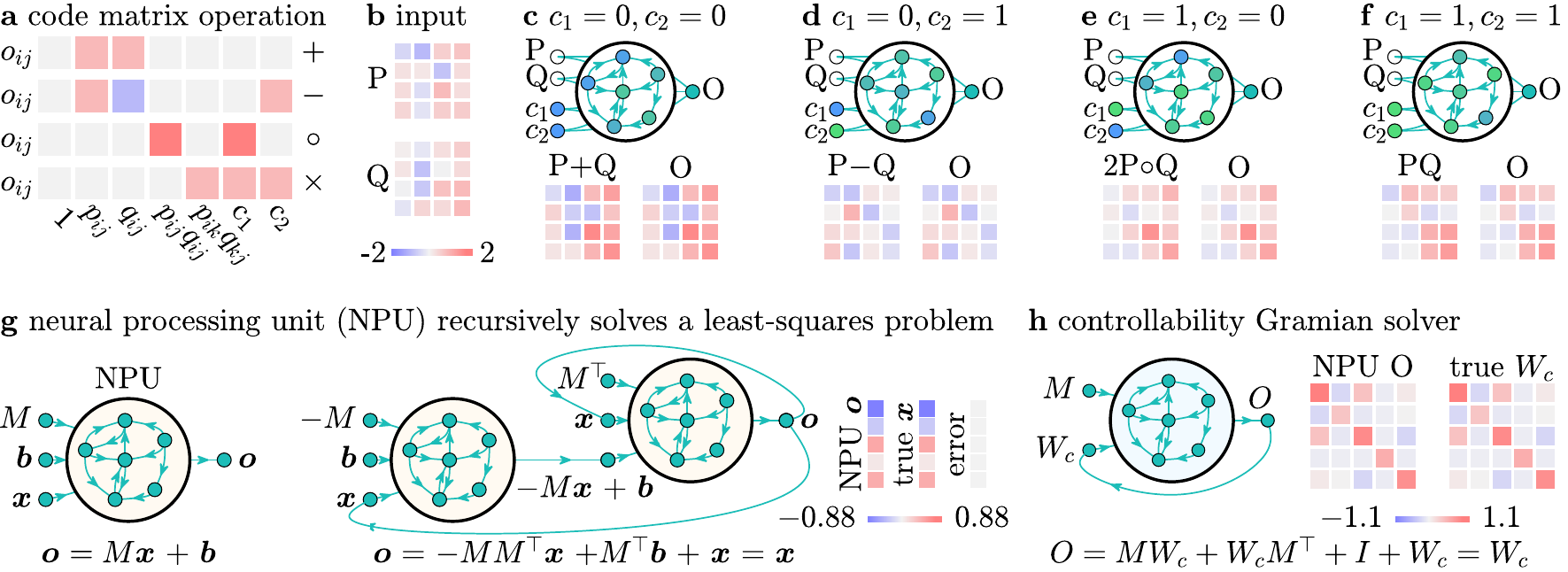}
	\caption{\label{fig:IO_matrix}\textbf{Programming matrix operations and solving equations through feedback.} (\textbf{a}) Required code for the RNN output to yield the matrix operations of addition, subtraction, element-wise multiplication, and matrix multiplication at various operating points determined by inputs $c_1$ and $c_2$. After programming these outputs $o_{ij}$ using a single matrix $W$, we input (\textbf{b}) two matrices P and Q to demonstrate that the output O of the RNN yields the (\textbf{c}) addition, (\textbf{d}) subtraction, (\textbf{e}) element-wise multiplication, and (\textbf{f}) matrix multiplication of P and Q, with indistinguishable error. (\textbf{g}) A neural processing unit (NPU) that outputs the affine transformation of its inputs. By chaining two NPUs, the output is the solution to the least-squares regression problem. Feeding the output back as inputs for $\bm{x}$ solves the least-squares problem, and (\textbf{h}) solves for the controllability Gramian by solving the Lyapunov equation. Typical relative error for all examples is 1\% or less.}
\end{figure}

This programming language defines two fundamentally new computing paradigms. First, it defines a natively \textit{symbolic} language. As opposed to modern-day silicon computers that are fundamentally numerical and require digitization into binary representations \cite{von1993first}, many computing tasks---from simple addition to complex matrix multiplication---are most naturally represented with variables. This representational mismatch requires complex algorithms and specialized hardware to numerically implement symbolic operations \cite{patterson2016computer}. Second, it defines natively \textit{distributed} processing. As opposed to modern-day silicon computers that must process information sequentially, many computing tasks---from linear regression to matrix equations---are not inherently defined sequentially, but rather through equivalence relations. 

We overcome these limitations with our recurrent neural programming language. Specifically, this language is natively symbolic, by which we mean that at the lowest level, neuronal activity is fundamentally a variable function of its inputs. Hence, we can define operations such as matrix addition and multiplication by programming an output matrix $W$ that selects the appropriate terms in the representational basis to form the output $\bm{o} = W\bm{r}$ (Fig.~\ref{fig:IO_matrix}a). We can even switch between different symbolic operations by programming our RNN about multiple operating points that we switch between using external inputs $c_1$ and $c_2$ (Fig.~\ref{fig:IO_matrix}b--f). 

Using these symbolic outputs, we provide two examples where we further solve symbolic equations in a natively distributed manner using feedback. In the first example, we use a simple neural processing unit (NPU) that is programmed to input a matrix $M$ and two vectors $\bm{b}$ and $\bm{x}$, and to output $\bm{o} = M\bm{x} + \bm{b}$. We chain multiple NPUs to produce more complex expressions---such as the least-squares solution to the linear regression $M\bm{x} = \bm{b}$---as the output. By feeding the output $\bm{o}$ back to drive the $\bm{x}$ inputs, the NPUs converge to the correct solution (Fig.~\ref{fig:IO_matrix}g). To understand this convergence, we notice that the output
\begin{equation}
\label{eq:map}
\bm{o}(\bm{x}) = -MM^\top\bm{x} + M^\top\bm{b} + \bm{x} = \bm{x},
\end{equation}
is a \textit{map} from $\bm{x}$ to $\bm{x}$. Through feedback, the RC evolves to the map's fixed point. In the second example, we calculate the \textit{controllability Gramian}, $W_c$, which is used extensively in many control applications from engineering \cite{pasqualetti2014controllability} to neuroscience \cite{karrer2020practical}. To compute $W_c$ on a neural computing architecture, we simply program the output to symbolically represent the equation that produces $W_c$, and feed the output back into the input for $W_c$. By evolving this feedback system, the RNN converges to the correct Gramian (Fig.~\ref{fig:IO_matrix}h).

\section{Dynamical Random Access Memory}
\begin{figure}[h!]
	\includegraphics[width=\textwidth]{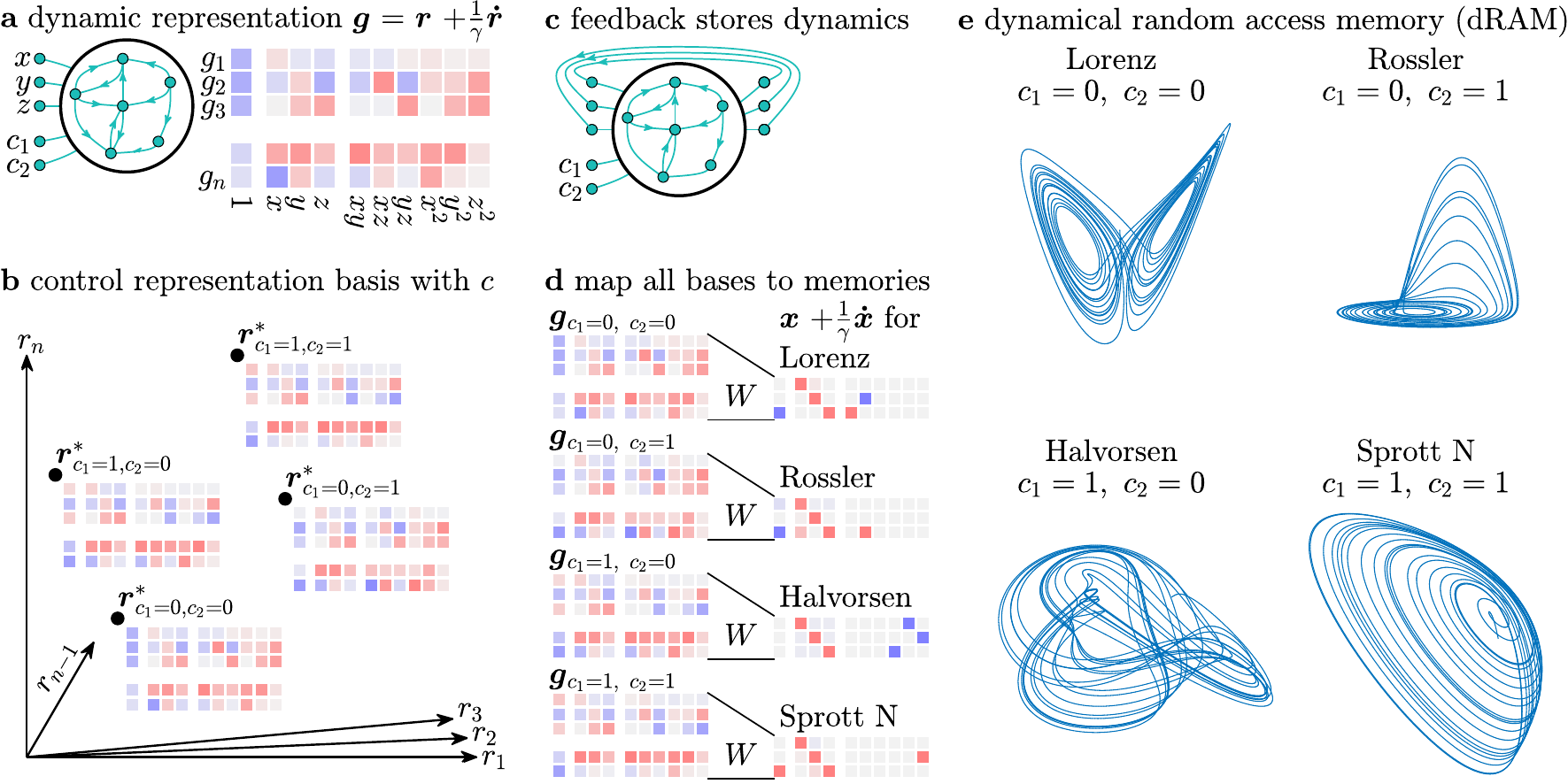}
	\caption{\label{fig:dRAM}\textbf{Programming dynamical memories at specific neural addresses.} (\textbf{a}) The dynamical representation of an RNN is the series expansion of $\bm{g}(A\bm{r} + B\bm{x} + \bm{d})$. By adding additional control input $c_1$ and $c_2$, (\textbf{b}) we can evaluate this representation at different equilibrium points of the RNN. (\textbf{c}) To store dynamical memories, we program a single output matrix $W$ that (\textbf{d}) maps each representation to a different dynamical attractor equation $\bm{x} + \frac{1}{\gamma}\dot{\bm{x}}$. (\textbf{e}) By changing the value of $c_1$ and $c_2$, we can retrieve the dynamical memories in the programmed locations, thereby forming our dynamical random access memory (dRAM).}
\end{figure}

In addition to defining fundamentally novel and natively symbolic computing paradigms, our recurrent neural programming language extends traditional concepts of computer memory from static to dynamic. Prior work in reservoir computing has demonstrated the power of training RNNs to store dynamical attractors as memories by copying exemplars \cite{sussillo2009generating}. By dynamical memory, we mean a time series that evolves according to a dynamical equation
\[\dot{\bm{x}} = \bm{f}(\bm{x}).
\]
Here we extend such work by programming dynamical memories without any exemplar time series, and by doing so at specific and addressable locations in the RNN's representation space, thereby yielding a dynamical Random Access Memory (dRAM).

To program dynamical memories, we must first extend our programming language because our current language only defines the RNN state $\bm{r}$ as a function of inputs $\bm{x}$, and encodes nothing about the evolution $\dot{\bm{r}}$ of the RNN. To encode dynamics, we rewrite Eq.~\ref{eq:RNN} as
\begin{equation}
\label{eq:RNN_rate}
\bm{g}(A\bm{r} + B\bm{x} + \bm{d}) = \bm{r} + \frac{1}{\gamma}\dot{\bm{r}}. 
\end{equation}
We then substitute the expression of $\bm{r}$ as a function of $\bm{x}$ from Eq.~\ref{eq:RNN_representation} to obtain $\bm{g}(A\bm{r}+B\bm{x}+\bm{d})$ as a purely linear combination of polynomial powers of $\bm{x}$ and its time derivatives (Fig.~\ref{fig:dRAM}a), which yields the final dynamical variant of our programming language. By evaluating Eq.~\ref{eq:RNN_rate} at different constant inputs $c_1$ and $c_2$, the representational basis also changes (Fig.~\ref{fig:dRAM},b).

The program then is a matrix $W$ that simultaneously copies the input state $W\bm{r} = \bm{x}$, and also copies the rate of change of the input state $W\dot{\bm{r}} = \dot{\bm{x}}$ such that
\begin{equation}
\label{eq:training_rate}
W\left(\bm{r} + \frac{1}{\gamma}\dot{\bm{r}}\right) = \bm{x} + \frac{1}{\gamma}\dot{\bm{x}}. 
\end{equation}
This feedback matrix produces an output $\bm{o} = W\bm{r}$, which---when used in place of the inputs $\bm{x}$ to drive the RC (Fig.~\ref{fig:dRAM}c)---will store the dynamical memory inside of the RNN. For example, we program $W$ to map the four representational bases generated by setting the control inputs $c_1$ and $c_2$ to four different dynamical memories (Fig.~\ref{fig:dRAM}d). Then, depending on the specific value of the control inputs, the output of the RNN yields the four different programmed dynamical memories (Fig.~\ref{fig:dRAM}e).

\section{A Recurrent Neural Virtual Machine}
\begin{figure}[h!]
	\includegraphics[width=\textwidth]{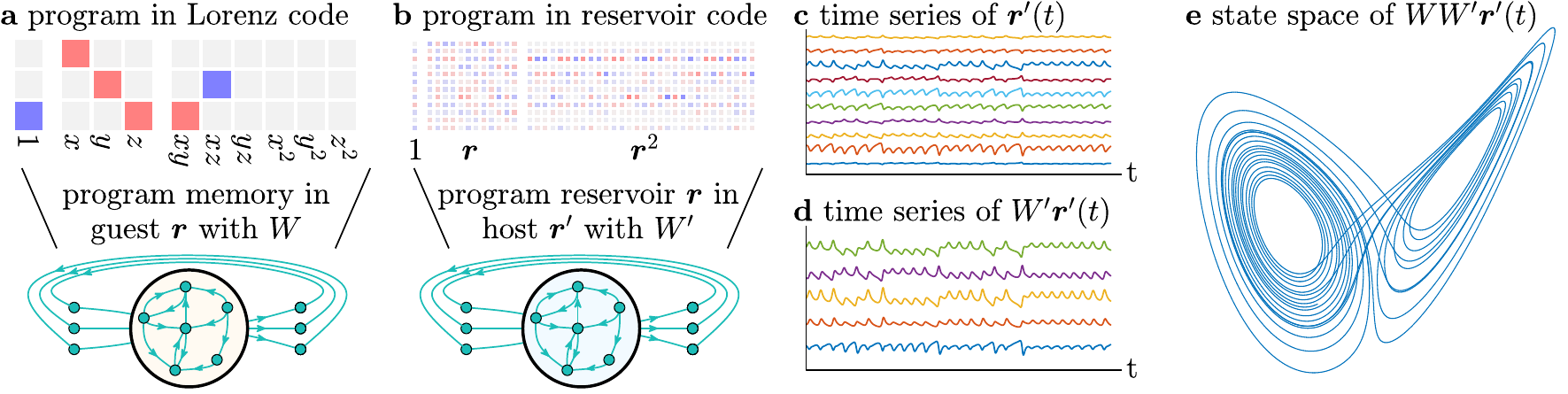}
	\caption{\label{fig:virtualize}\textbf{Programming an RNN inside of another RNN.} (\textbf{a}) The guest RNN is programmed to store a chaotic Lorenz attractor as a memory using the feedback matrix $W$. The code is $\bm{x} + \frac{1}{\gamma}\dot{\bm{x}}$ according to Eq.~\ref{eq:training_rate}. (\textbf{b}) The host RNN is programmed to store the guest RNN as a memory using feedback matrix $W'$. (\textbf{c}) The host RNN is evolved autonomously forward in time as $\bm{r}'(t)$, which (\textbf{d}) simulates the dynamics of the guest RNN as $W'\bm{r}'(t)$, which (\textbf{e}) simulates the dynamics of the chaotic Lorenz attractor as $WW'\bm{r}'(t)$.}
\end{figure}

This ability to store dynamical memories allows us to concretely implement core ideas from computer software into RNNs. Specifically, because an RNN can be programmed to evolve about a dynamical system, and an RNN itself is a dynamical system, we arrive at a curious question: can an RNN be programmed to simulate another RNN? This concept is referred to as a \textit{virtual machine} \cite{rosenblum2005virtual}, where a host computer simulates the hardware of a guest computer. We demonstrate that the answer to this question is yes.

First, we program the guest RNN $\bm{r}$ to store a Lorenz attractor as a dynamical memory using a feedback matrix $W$ (Fig.~\ref{fig:virtualize}a). Then, we write out the code for this programmed guest RNN, and program a separate host RNN $\bm{r}'$ to store the programmed guest RNN as a dynamical memory using a feedback matrix $W'$ (Fig.~\ref{fig:virtualize}b). Finally, when we evolve the host RNN (Fig.~\ref{fig:virtualize}c), we find that it simulates the dynamics of the guest RNN through $W'\bm{r}$ (Fig.~\ref{fig:virtualize}d), which has stored the programmed Lorenz attractor through $WW'\bm{r}$ (Fig.~\ref{fig:virtualize}e).

\section{A Logical Calculus Using Recurrent Neural Circuits}
\begin{figure}[h!]
	\includegraphics[width=\textwidth]{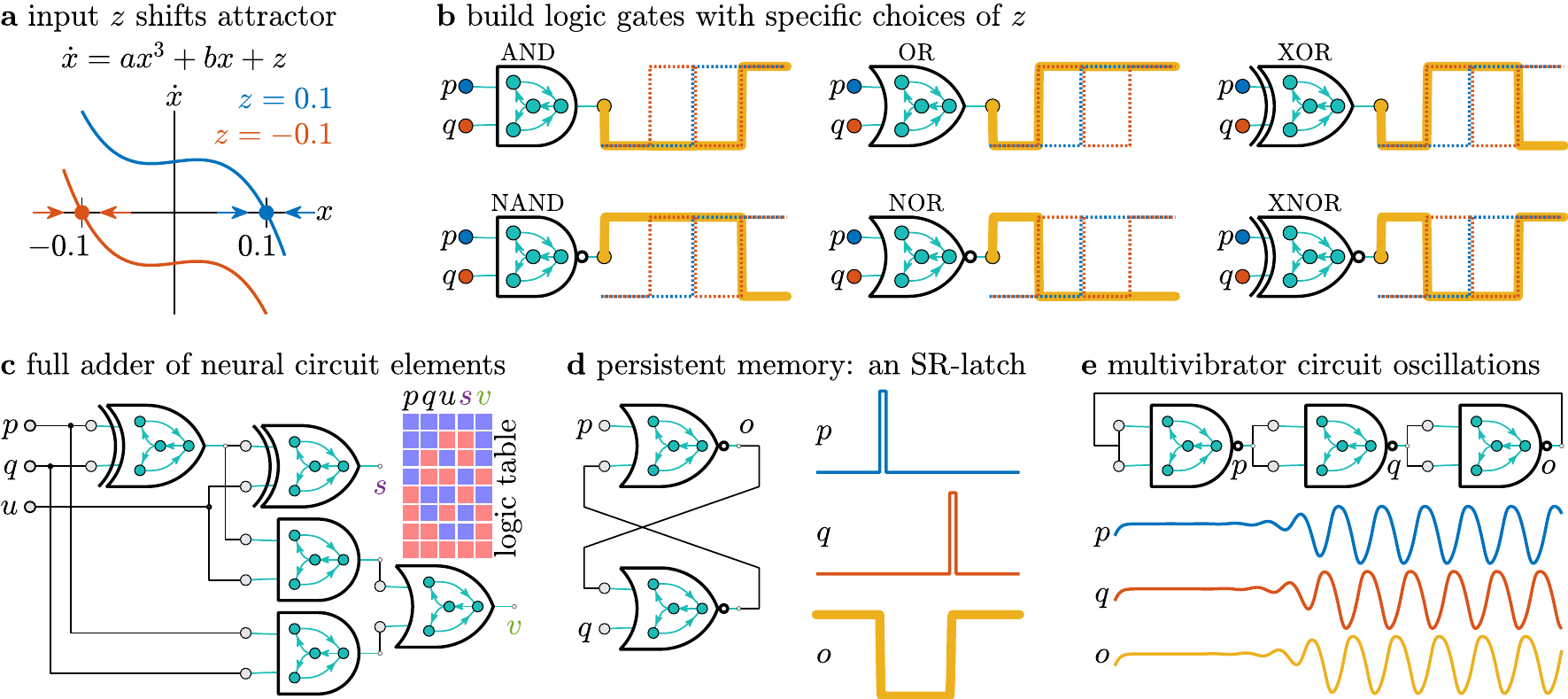}
	\caption{\label{fig:neural_circuits}\textbf{Programming logic gates and circuits using dynamical neural networks.} (\textbf{a}) Phase diagram of a cubic dynamical system. When $z = 0.1$, the variable $x$ tends towards the stable fixed point $x^* = 0.1$. When $z = -0.1$, the system bifurcates, and a new stable fixed point emerges at $x^* = -0.1$. (\textbf{b}) By setting $z$ equal to various products of two input variables $p$ and $q$, the output evolves according to different Boolean logic gates, and we program these logic gate dynamics into our RNNs. By connecting these neural logic gates, we can form neural circuits that (\textbf{c}) add Boolean numbers, (\textbf{d}) store persistent Boolean states according to a set-reset (SR) latch, and (\textbf{e}) oscillate at a fixed phase difference due to the propagation delay of inversion operations.}
\end{figure}

This dynamical programming language allows us to greatly expand the computational capability of our RNNs by programming neural implementations of logic gates. While prior work has established the ability of biological and artificial networks to perform computations, here we provide an implementation that makes full use of existing computing frameworks. We program logic gates into distributed RNNs by using a simple dynamical system
\begin{equation}
\label{eq:cubic}
\dot{x} = ax^3 + bx + z,
\end{equation}
where $a, b,$ and $z$ are parameters. This particular system has the nice property of \textit{hysteresis}, where when $z = 0.1$, the value of $x$ converges to $x = 0.1$, but when $z = -0.1$, the value of $x$ jumps discontinuously to converge at $x = -0.1$ (Fig.~\ref{fig:neural_circuits}a). This property enables us to program logic gates (Fig.~\ref{fig:neural_circuits}b). Specifically, by defining the variable $z$ as a product of two input variables $p$ and $q$, we can program in the dynamics in Eq.~\ref{eq:cubic} to evolve to $-0.1$ or $0.1$ for different patterns of $p$ and $q$. 

These logic gates can now take full advantage of existing computing frameworks. For example, we can construct a full adder using neural circuits that take Boolean values $p$ and $q$ as the two numbers to be added, and a ``carry'' value from the previous addition operation. The adder outputs the sum $s$ and the output carry $v$. We show the inputs and outputs of a fully neural adder in Figure~\ref{fig:neural_circuits}c, forming the basis of our ability to program neural logic units (NLU), which are neural analogs of existing arithmetic logic units (ALU). 

The emulation of these neural logic gates to circuit design extends even to recurrent circuit architectures. For example, the set-reset (SR) latch---commonly referred to as a flip-flop---is a circuit that holds persistent memory, and is used extensively in computer RAM. We construct a neural SR-latch using two NOR gates with two inputs, $p$ and $q$ (Fig~\ref{fig:neural_circuits}d). When $p=0.1$ is pulsed high, the output $o = -0.1$ changes to low. When $q$ is pulsed high, the output changes to high. When both $p$ and $q$ are held low, then the output is fixed at its most recent value (Fig~\ref{fig:neural_circuits}d). As another example, we can chain an odd number of inverting gates (i.e., \textsc{nand, nor}, and \textsc{xor}) to construct a multivibrator circuit that generates oscillations (Fig.~\ref{fig:neural_circuits}e). Because the output of each gate will be the inverse of its input, if $p$ is high, then $q$ is low, and $o$ is high. However, if we use $o$ as the input to the first gate, then $p$ must switch to low. This discrepancy produces constant fluctuations in the states of $p, q,$ and $o$, which generate oscillations that are offset by the same phase (Fig.~\ref{fig:neural_circuits}e).

\section{Game Development on Recurrent Neural Architectures}
\begin{figure}[h!]
	\includegraphics[width=\textwidth]{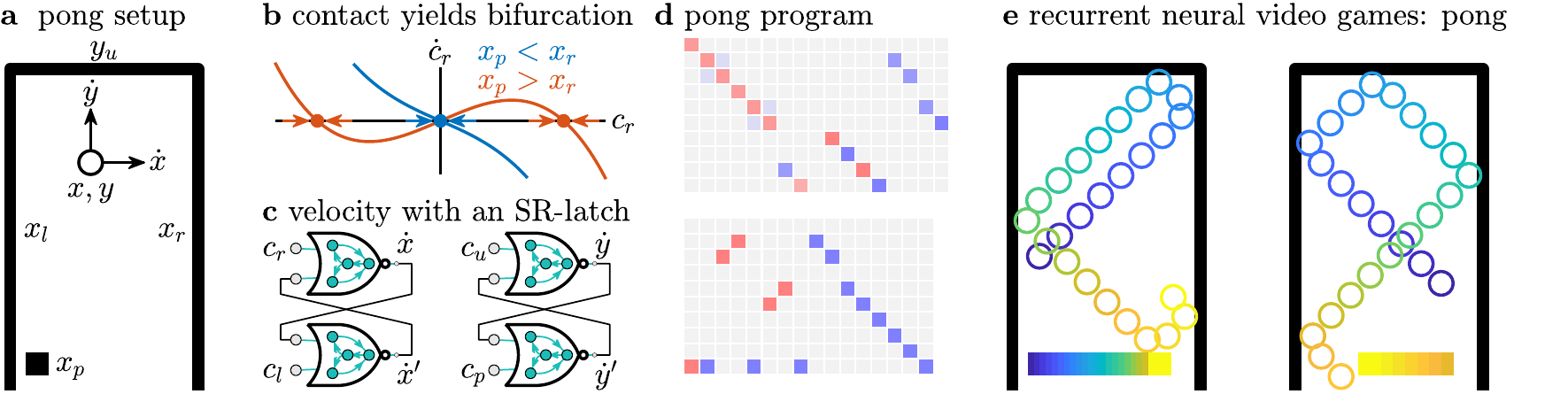}
	\caption{\label{fig:pong}\textbf{Programming pong using neural circuits and bifurcations.} (\textbf{a}) Design of a pong variant. The wall positions ($x_l, x_r, y_u$) and the paddle's y-coordinate are fixed as constants. The variables are the ball's position ($x,y$) and velocity ($\dot{x},\dot{y}$), the paddle's position ($x_p$), and the variables determining contact with the walls and paddle ($c_l, c_r, c_u, c_p$). (\textbf{b}) Contact detection with the right wall is implemented using a supercritical pitchfork bifurcation by scaling the $b$ term in Eq.~\ref{eq:cubic} by $x - x_r$. When $x < x_r$, the contact variable $c_r$ goes to 0. When $x > x_r$, a bifurcation occurs and $c_r$ becomes non-zero. (\textbf{c}) These contact variables are used to drive an SR-latch whose output is the ball's velocity. (\textbf{d}) Code for the dynamical pong game. (\textbf{e}) An RNN simulating a game of pong in its head. The color from blue to yellow represents the evolution of time. The square on the bottom is the movement of the paddle, and the circle is the movement of the marker.}
\end{figure}

To demonstrate the flexibility and capacity of our framework, we program a variant of the game ``pong'' into our RNN as a dynamical system. We begin with the game design by defining the relevant variables and behaviors (Fig.~\ref{fig:pong}a). The variables are the coordinates of the ball, $x,y$, the velocity of the ball, $\dot{x}, \dot{y}$, and the position of the paddle, $x_p$. Additionally, we have the variables that determine contact with the left, right, and upper walls as $c_l, c_r$, and $c_u$, respectively, and the variable that determines contact with the paddle, $c_p$. The behavior that we want is for the ball to travel with a constant velocity until it hits either a wall or the paddle, at which point the ball should reverse direction.

Here, we run into our first problem: how do we represent contact detection---a fundamentally discontinuous behavior---using continuous dynamics? Recall that we have already done so to program logic gates in Fig.~\ref{fig:neural_circuits}a by using the bifurcation of the cubic dynamical system in Eq.~\ref{eq:cubic}. Here, we will use the exact same equation, except rather than changing the parameter $z$ to shift the dynamics up and down (Fig.~\ref{fig:neural_circuits}a), we will set the parameter $b$ to skew the shape. As an example, for the right-wall contact $c_r$, we will let $b = x - x_r$ (Fig.~\ref{fig:pong}b). When the ball is to the left such that $x < x_r$, then $c_r$ approaches 0. When the ball is to the right such that $x > x_r$, then $c_r$ becomes non-zero.

To set the velocity of the ball, we use the SR-latch developed in Fig.~\ref{fig:neural_circuits}d. When neither wall is in contact, then $c_r$ and $c_l$ are both low, and the latch's output does not change. When either the right or the left wall is in contact, then either $c_r$ or $c_l$ pulses the latch, producing a shift in the velocity (Fig.~\ref{fig:pong}c). Combining these dynamical equations together produces the code for our pong program (Fig.~\ref{fig:pong}d), and the time-evolution of our programmed RNN simulates a game of pong (Fig.~\ref{fig:pong}e).

\section{Discussion}
Neural computation exists simultaneously at the core of and the intersection between many fields of study. From the differential binding of neurexins in molecular biology \cite{sudhof2017synaptic} and neural circuits in neuroscience \cite{lerner2016communication,feller1999spontaneous,calhoon2015resolving,maass2007computational,clarke2013emerging}, to the RNNs in dynamical systems \cite{sussillo2014neural} and neural replicas of computer architectures in machine learning \cite{graves2016hybrid}, the analogy between neural and silicon computers has generated growing and interdisciplinary interest. Our work provides one possible realization of this analogy by defining a dynamical programming language for RNNs that takes full advantage of their natively continuous and symbolic representation, their parallel and distributed processing, and their dynamical evolution. 

This work also makes an important contribution to the increasing interest in alternative computation. A clear example is the vast array of systems---such as molecular \cite{kompa2001molecular}, DNA \cite{zhang2012reversible}, and single photon \cite{pittman2003experimental}---that implement Boolean logic gates. Other examples include the design of materials that compute \cite{fang2016pattern,stern2021supervised} and store memories \cite{pashine2019directed,chen2021reprogrammable}. Perhaps of most relevance are physical instantiations of reservoir computing in electronic, photonic, mechanical, and biological systems \cite{tanaka2019recent}. Our work demonstrates the potential of alternative computing frameworks to be fully programmable, thereby shifting paradigms away from imitating silicon computer hardware, and towards defining native programming languages that bring out the full computational capability of each system.

One of the main current limitations is the linear approximation of the RC dynamics. While prior work demonstrates significant computational ability for RCs with largely fluctuating dynamics (i.e., computation at the edge of chaos \cite{boedecker2012information}), the approximation used in this work requires that the RC states stay reasonably close to the operating points. While we are able to program a single RC at multiple operating points that are far apart, the linearization is a prominent limitation. Future extensions would use more advanced dynamical approximations into the bilinear regime using Volterra kernels \cite{svoronos1980bilinear} or Koopman composition operators \cite{bevanda2021koopman} to better capture nonlinear behaviors. 

Finally, we report in the Supplementary Section V an analysis of the gender and racial makeup of the authors we cited in a Citation Diversity Statement.

\section{Data \& Code Availability Statement}
There is no data with mandated deposition used in the manuscript or supplement. All analysis and figures were created programmatically in MATLAB, and the code will be made public upon acceptance of the manuscript.

\section{Acknowledgments}
\noindent We gratefully acknowledge Dr. Melody X. Lim, Dr. Kieran A. Murphy, Harang Ju, Dale Zhou, and Dr. Jeni Stiso for conversations and comments on the manuscript. JZK acknowledges support from the National Science Foundation Graduate Research Fellowship No. DGE-1321851. DSB acknowledges support from the John D. and Catherine T. MacArthur Foundation, the ISI Foundation, the Alfred P. Sloan Foundation, an NSF CAREER award PHY-1554488, and from the NSF through the University of Pennsylvania Materials Research Science and Engineering Center (MRSEC) DMR-1720530.

\end{document}


\bibliographystyle{naturemag}
\title{Supplement to ``A Neural Programming Language for the Reservoir Computer''}
\author{Jason Z. Kim}
\affiliation{Department of Bioengineering, University of Pennsylvania, Philadelphia, PA, 19104}
\author{Dani S. Bassett}
\affiliation{Departments of Bioengineering, Physics \& Astronomy, Electrical \& Systems Engineering, Neurology, and Psychiatry, University of Pennsylvania, Philadelphia, PA, 19104}
\affiliation{Santa Fe Institute, Santa Fe, NM 87501}
\affiliation{To whom correspondence should be addressed: dsb@seas.upenn.edu}
\date{\today}
\maketitle

\section{Deriving the expansion of reservoir state}
In the main text, we provide the outline of a derivation for the neural programming language. Here, we expand upon this derivation in more detail. As in the main text, we begin with the reservoir equation
\begin{equation}
\label{eq:rnn}
\frac{1}{\gamma}\dot{\bm{r}}(t) = -\bm{r}(t) + \bm{g}(A\bm{r}(t) + B\bm{x}(t) + \bm{d}). 
\end{equation}
As in the main text, we will omit the explicit denotation of the time variable using $(t)$ for conciseness and clarity. Our goal is to write $\bm{r}$ as an explicit and symbolic function of $\bm{x}$, which is difficult to do for nonlinear systems. Hence, the first step is to linearize this equation about a static operating point $\bm{r}^*$, which yields
\begin{equation*}
\frac{1}{\gamma}\dot{\bm{r}} = -\bm{r} + \bm{g}(A\bm{r}^* + B\bm{x} + \bm{d}) + \D\bm{g}(A\bm{r}^* + B\bm{x} + \bm{d}) \circ (A\bm{r} - A\bm{r}^*), 
\end{equation*}
where $\circ$ is the element-wise product operation. Rearranging this equation to group the terms with the $\bm{r}$ variable yields
\begin{equation*}
\frac{1}{\gamma} \dot{\bm{r}} = -\bm{r} + \underbrace{\D\bm{g}(A\bm{r}^* + B\bm{x} + \bm{d})\circ (A\bm{r})}_{\mathrm{bilinear}} + \bm{g}(A\bm{r}^* + B\bm{x} + \bm{d}) - \D\bm{g}(A\bm{r}^* + B\bm{x} + \bm{d}) \circ (A\bm{r}^*). 
\end{equation*}
However, we notice a \textit{bilinear} term in this expansion that contains both $\bm{x}$ and $\bm{r}$. Hence, the second step in this approximation is to linearize the bilinear term by evaluating $\bm{x}$ at an operating point $\bm{x}^*$ to yield
\begin{equation*}
\frac{1}{\gamma} \dot{\bm{r}} = \underbrace{(\D\bm{g}(A\bm{r}^* + B\bm{x}^* + \bm{d}) \circ A - I)}_{A^*}\bm{r} + \bm{g}(A\bm{r}^* + B\bm{x} + \bm{d}) - \D\bm{g}(A\bm{r}^* + B\bm{x} + \bm{d}) \circ (A\bm{r}^*). 
\end{equation*}
Rewriting this equation using the additional substitution of $\bm{d}^* = A\bm{r}^* + \bm{d}$ yields
\begin{equation}
\label{eq:rnn_linear}
\frac{1}{\gamma}\dot{\bm{r}} = A^*\bm{r} + \underbrace{\bm{g}(B\bm{x} + \bm{d}^*) - \D\bm{g}(B\bm{x}+\bm{d}^*) \circ (A\bm{r}^*)}_{\bm{u}(\bm{x}(t))}.
\end{equation}
The term in the underbrace---$\bm{u}(\bm{x}(t))$---is precisely the effective input in the main text. This equation is now a linear system of equations whose states we can write as the convolution
\begin{equation*}
\bm{r}(t) = \gamma \int_{-\infty}^t e^{\gamma A^*(t-\tau)}\bm{u}(\bm{x}(\tau))\D\tau. 
\end{equation*}
Substituting the expression for the effective input yields
\begin{equation}
\bm{r}(t) = \gamma \int_{-\infty}^{t} e^{\gamma A^*(t-\tau)} \bm{g}(B\bm{x}(\tau) + \bm{d}^*) \D\tau - \gamma \int_{-\infty}^{t} e^{\gamma A^*(t-\tau)}A\bm{r}^* \circ \D\bm{g}(B\bm{x}(\tau)+\bm{d}^*) \D\tau.
\end{equation}
While this convolution is technically a symbolic function of $\bm{x}$, we would like a simpler algebraic expression that does not involve integrals. Hence, we evaluate the integrals by performing a series expansion of $\bm{g}$ and $\D\bm{g}$ with respect to $\tau = t$ to yield
\begin{equation}
\label{eq:time_expand}
\begin{split}
\bm{g}(B\bm{x}(\tau)+\bm{d}^*) &\approx \bm{g}(B\bm{x}(t) + \bm{d}^*) - (t-\tau)\D\bm{g}(B\bm{x}(t)+\bm{d}^*)\circ (B\dot{\bm{x}}),\\
\D\bm{g}(B\bm{x}(\tau)+\bm{d}^*) &\approx \D\bm{g}(B\bm{x}(t) + \bm{d}^*) - (t-\tau)\D^2\bm{g}(B\bm{x}(t)+\bm{d}^*)\circ (B\dot{\bm{x}}).
\end{split}
\end{equation}
Using the following integration identities:
\begin{equation*}
\begin{split}
&\int_{-\infty}^{t} e^{A(t-\tau)}\D\tau = -A^{-1}\\
&\int_{-\infty}^{t} e^{A(t-\tau)}(t-\tau)\D\tau = A^{-2},
\end{split}
\end{equation*}
we can evaluate the integrals to arrive at an algebraic expression
\begin{equation}
\label{eq:basis}
\begin{split}
\bm{r}(t) = &A^{*-1} (\D\bm{g}(B\bm{x}+\bm{d}^*)\circ(A\bm{r}^*) - \bm{g}(B\bm{x} + \bm{d}^*)) +\\
&\frac{1}{\gamma}A^{*-2}\left((\D^2\bm{g}(B\bm{x}+\bm{d}^*) \circ (A\bm{r}^*) - \D\bm{g}(B\bm{x}+\bm{d}^*)) \circ (B\dot{\bm{x}})\right).
\end{split}
\end{equation}
The right-hand side of this equation is precisely the analytic function $\bm{h}$ in the main text. Finally, by taking the Taylor series expansion of Eq.~\ref{eq:basis} with respect to $\bm{x}$, we obtain an algebraic expression of the reservoir neuron states, $\bm{r}(t)$, as a weighted sum of polynomials in $\bm{x}$ and $\dot{\bm{x}}$. By taking additional, higher-order terms in the expansion of Eq.~\ref{eq:time_expand}, we resolve additional time derivatives $\ddot{\bm{x}}, \dddot{\bm{x}}, \dotsm$ in the expansion, yielding the analytic function $\bm{h}(\bm{x},\dot{\bm{x}},\ddot{\bm{x}},\dotsm)$ in the main text.
\newpage

\section{Upper bound on the goodness of the linearization}
Given the number of steps that we took in deriving the reservoir state as an algebraic function of inputs $\bm{x}$ and their time derivatives, it is important to quantify the goodness of the approximation. The first and most fundamental step in the approximation process was the linearization of the system to obtain Eq.~\ref{eq:rnn_linear}. Hence, we will first compare the trajectory $\bm{r}(t)$ of the full nonlinear equation using Eq.~\ref{eq:rnn} to the trajectory $\bm{r}'(t)$ of the linearized equation using Eq.~\ref{eq:rnn_linear}. To systematically study the approximation, we compute the relative error given by the norm of $\bm{r}(t) - \bm{r}'(t)$ divided by the norm of $\bm{r}(t) - \bm{r}^*$, where $\bm{r}^*$ is the operating point of the system. 

We vary two important parameters that determine the goodness of the approximation: the magnitude of $A$ and the magnitude of $B$. Specifically, we draw the entries of $A$ and $B$ from uniform, independent, and random distributions ranging from $-1$ to $1$, and then linearly scale $A$ by a scalar $a$, and $B$ by a scalar $b$. We fix the RC to have $n=1000$ neurons, and we require $A$ to be 5\% dense and $B$ to be fully dense. We also fix $\gamma = 100$, and we draw the operating point $\bm{r}^*$ from a uniform random distribution ranging from $-0.1$ to $0.1$. 

\begin{figure}[h!]
	\includegraphics[width=\textwidth]{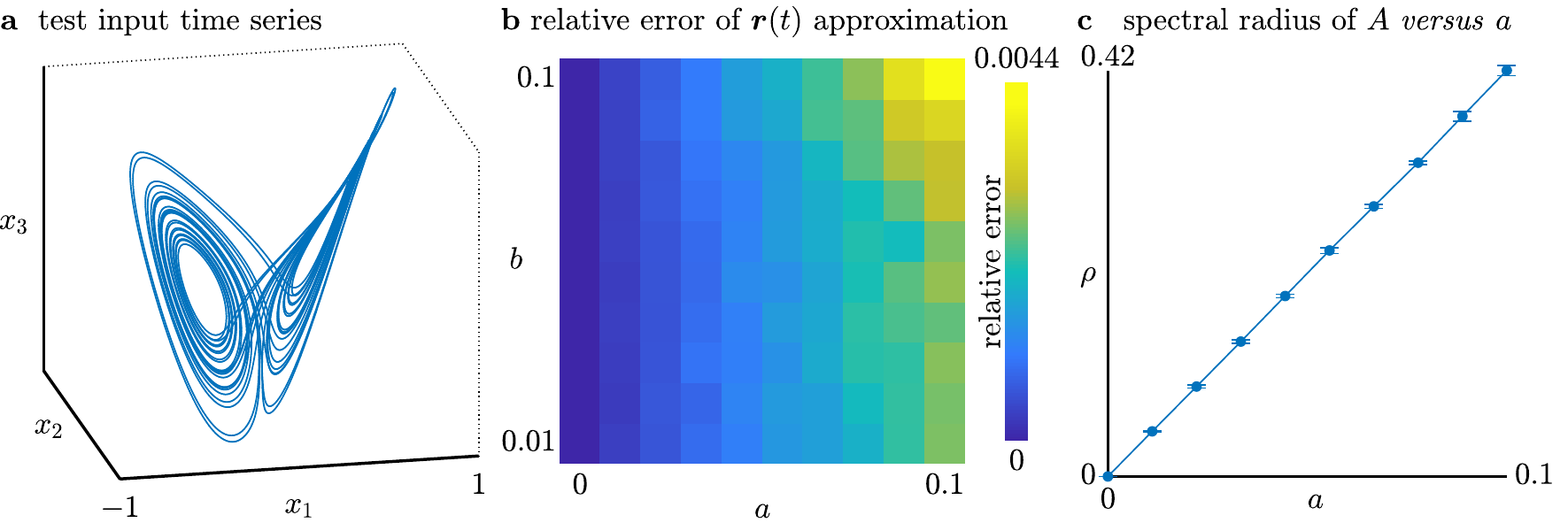}
	\caption{\label{suppfig:test_approx_lin}\textbf{Testing the goodness of the linearization.} (\textbf{a}) Chaotic Lorenz attractor which is the test input into our RNN. All axes range from $-1$ to 1. (\textbf{b}) Relative error of the reservoir states between those generated using the full nonlinear expression, $\bm{r}(t)$ (Eq.~\ref{eq:rnn}), and those generated using the linearized expression, $\bm{r}'(t)$ (Eq.~\ref{eq:rnn_linear}). We notice the errors are less than 1\%. (\textbf{c}) Plot of the means and standard deviations of the spectral radius of the adjacency matrix $A$ given by the magnitude of its largest eigenvalue $\rho$. Each point is the average $\rho$ across the 10 different instantiations of $A$ for each value of $a$.}
\end{figure}

As our test input, we will use a time series generated from the 3-dimensional Lorenz attractor that has been shifted to be centered around $\bm{x} \approx \bm{0}$ and scaled such that its states evolve roughly between $-1 \leq x_1,x_2,x_3 \leq 1$ using the equations
\begin{equation}
\label{eq:lorenz}
\begin{split}
\dot{x}_1 &= \sigma(x_2 - x_1)\\
\dot{x}_2 &= x_1(\rho - (20x_3+27)) - x_2\\
\dot{x}_3 &= 20x_1x_2 - \beta \left(x_3+\frac{27}{20}\right),
\end{split}
\end{equation}
where $\sigma = 10$, $\rho = 28$, and $\beta = 8/3$. The trajectory used for training is shown in Fig.~\ref{suppfig:test_approx_lin}a. As can be seen, the relative error increases with larger scaling values of $a$ and $b$, but the error remains solidly below 1\% for this wide range of parameters (Fig.~\ref{suppfig:test_approx_lin}b). 

To contextualize these scaling values, we first note that any dependence of the approximation on $b$ can be addressed simply by rescaling the input. For example, if we were to rescale the input time series in Fig.~\ref{suppfig:test_approx_lin}a to one tenth of its current magnitude, then the relative error in Fig.~\ref{suppfig:test_approx_lin}b at $b = 0.1$ would significantly decrease to the relative error at $b = 0.01$. This is because mathematically, $B$ and $\bm{x}$ are multiplied, such that rescaling the input can always provide smaller relative error.

To further contextualize this result, we next note that the behavior of the RC is intimately tied to both the scaling $a$ and the number of neurons $n$. Specifically, the dynamical behavior of the RC depends fundamentally on the \textit{spectral radius} $\rho$ of the adjacency matrix, which is given by the magnitude of its largest eigenvalue. We show that in our $n=1000$ neuron network, the largest scaling value corresponds to a spectral radius of $\rho \approx 0.42$, thereby demonstrating that our approximation accurately captures the non-trivial internal dynamics of the RC (Fig.~\ref{suppfig:test_approx_lin}c).

In sum, we demonstrate that a naive test of the linearization accuracy yields a good and robust approximation across a typical range of parameters. Further, this accuracy can be improved by rescaling the magnitude of the input---a common practice in many disciplines---as well as the selection of a small spectral radius.
\newpage

\section{Goodness of the expansions in input variables and time}
Now that we have a clear idea of how our parameter choices affect the accuracy of the linearization, we next quantify the accuracy of the subsequent approximations in time (Eq.~\ref{eq:time_expand}) and in the input variables (Eq.~\ref{eq:basis}. To do this, we perform the same protocol as in the previous section by driving the RNN according to the chaotic Lorenz attractor (Eq.~\ref{eq:lorenz}). We measure the relative error between the reservoir states generated by the full nonlinear equation in Eq.~\ref{eq:rnn}, $\bm{r}(t)$, and those generated by the polynomial expansion of Eq.~\ref{eq:basis}, $\bm{r}^\dagger(t)$, as a function of the number of expansions in time (i.e., more time derivatives) and in input variables (i.e., higher-order powers of $\bm{x}$). We use the same $n, \gamma$, and $\bm{r}^*$ distribution.

\begin{figure}[h!]
	\includegraphics[width=\textwidth]{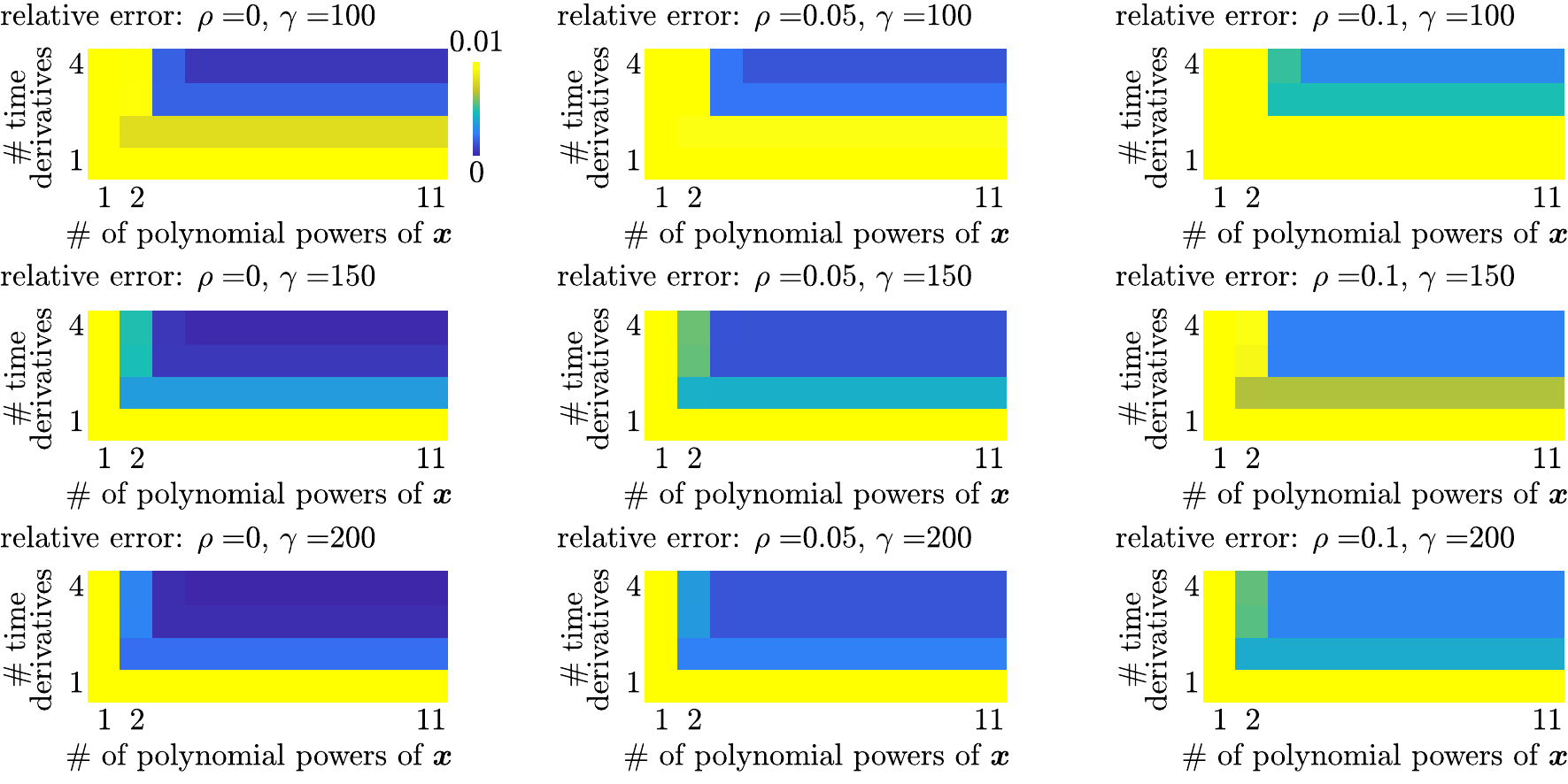}
	\caption{\label{suppfig:test_approx_expand}\textbf{Goodness of the expansion.} Relative error between the true RC time series (Eq.~\ref{eq:rnn}) and the estimated RC time series (Eq.~\ref{eq:basis}) at various spectral radii $\rho$ and time constants $\gamma$. Generally, the approximation improves with smaller $\rho$, larger $\gamma$, and more expansion terms in the number of time derivatives and the number of polynomial powers of $\bm{x}$.}
\end{figure}

We do not require too many time derivatives or polynomial powers of $\bm{x}$ to achieve a relative error that is significantly lower than $0.01$ (Fig.~\ref{suppfig:test_approx_expand}), and the approximation converges rapidly with smaller spectral radius $\rho$ and larger $\gamma$. Similar to the previous section, what matters is not the absolute value of $\gamma$, but rather the relative ratio of $\gamma$ to the time derivatives $\dot{\bm{x}}, \ddot{\bm{x}}, \dotsm$, and we can improve the accuracy simply by changing the time scale of $\bm{x}(t)$. 
\newpage

\section{Simulation parameters}
Throughout the main text, we provide many examples involving different dynamical equations and RC parameters. Here we explicitly write out the parameter choices and equations used. In line with the previous sections on the goodness of the approximation, the connectivity matrix $A$ is 0.05\% dense, whereas the input matrix $B$ is fully dense. The elements of $A$ and $B$ are always drawn from uniform random distributions. We discuss $B$ with respect to the amount by which it is scaled, and $A$ with respect to its spectral radius $\rho$. All simulations were run by integrating the continuous-time differential equations using the 4th order Runge-Kutta method with a time step of $\D t = 0.001$. Hence, a simulation that is run for $T$ time \textit{units} means that the simulation was evolved by $T/\D t$ steps. Throughout, we used $\gamma = 100$, and drew $\bm{r}^*$ from a random uniform distribution whose values ranged between $-0.5$ and $0.5$.

\subsection{Parameters for ``Symbolic Decomposition of Neural Representation''}
In this rotation example, we used a reservoir with $n = 100$ neurons, $\rho = 0.01$, and $B$ is scaled by 0.1. For the Thomas attractor, we chose an initial condition of $x_1(0) = 0, x_2(0) = 0,$ and $x_3(0) = 1$, and then we evolved the Thomas attractor for 100 time units, drove the RC for those 100 time units, and discarded the first 20 time units to allow the RC to forget about its initial condition.

\subsection{Parameters for ``Natively Distributed and Symbolic Computation''}
In these symbolic computation examples in Fig.~2 of the main text, we used an RC with $n = 5000$ neurons, $\rho = 0$, and $B$ that was scaled by 0.0005. For panels a--f, we used one such RC. The elements of the $4 \times 4$ matrices $P$ and $Q$ were drawn randomly and uniformly between $-1$ and $1$. For panel g, the NPU has $n=2500$ neurons, and two NPUs are chained together to form a system of $n=5000$ neurons. The elements of the $5 \times 5$ matrix $M$ were drawn randomly and uniformly between $-1$ and $1$, and the elements of the $5 \times 1$ vector $\bm{b}$ were drawn randomly and uniformly between $-0.5$ and $0.5$. For panel $h$, the RC has $n = 5000$ neurons. The elements of the $5 \times 5$ matrix $M$ were drawn randomly and uniformly between $-0.5$ and $-0.5$, and then we subsequently subtracted the identity matrix from $M$. This subtraction is to ensure that $M$ is stable---all real eigenvalues are negative---because the Lyapunov equation does not possess a solution otherwise.

\subsection{Parameters for ``Dynamical Random Access Memory''}
In programming the RC of Fig.~3 of the main text, we used an RC with $n = 400$ neurons, $\rho = 0.01$, and $B$ that was scaled by 0.05. The equations for the dynamical attractors are
\begin{equation*}
\begin{split}
\dot{x}_1 &= 10(x_2-x_1)\\
\dot{x}_2 &= x_1 - x_2 - 20x_1x_3\\
\dot{x}_3 &= 20x_1x_2 - \frac{8}{3}\left(x_3 - \frac{27}{20}\right),
\end{split}
\end{equation*}
for the Lorenz, 
\begin{equation*}
\begin{split}
\dot{x}_1 &= -5_2 - 5x_3 - 3\\
\dot{x}_2 &=  5x_1 + x_2\\
\dot{x}_3 &=  \frac{250}{3}x_1x_3 + 50x_1 - 28.5x_3 - 17.04,
\end{split}
\end{equation*}
for the Rossler,
\begin{equation*}
\begin{split}
\dot{x}_1 &= -2.1x_1 - 6x_2 - 6x_3 - 15x_2^2\\
\dot{x}_2 &= -2.1x_2 - 6x_3 - 6x_1 - 15x_3^2\\
\dot{x}_3 &= -2.1x_3 - 6x_1 - 6x_2 - 15x_1^2,
\end{split}
\end{equation*}
for the Halvorsen, and
\begin{equation*}
\begin{split}
\dot{x}_1 &= -8x_2\\
\dot{x}_2 &= \frac{25}{4}x_1 + 5x_3^2\\
\dot{x}_3 &= \frac{5}{4} + 20x_2 - 10x_3,
\end{split}
\end{equation*}
for the Sprott N.

\subsection{Parameters for ``A Recurrent Neural Virtual Machine''}
In programming the RC of Fig.~4 in the main text, we used a guest RC with $n = 12$ neurons, and a host RC with $n = 1000$ neurons. For both the guest and host RCs, $\rho = 0$, and $B$ was scaled by 0.1. The Lorenz equation that was programmed was
\begin{equation*}
\begin{split}
\dot{x}_1 &= x_2-x_1\\
\dot{x}_2 &= \frac{1}{10}x_1 - \frac{1}{10}x_2 - 20x_1x_3\\
\dot{x}_3 &= 20x_1x_2 -\frac{4}{15}x_3 - 0.036.
\end{split}
\end{equation*}

\subsection{Parameters for ``A Logical Calculus Using Recurrent Neural Circuits''}
In programming the RCs of Fig.~5 in the main text, each logic gate used an RC with $n = 30$ neurons, with $\rho = 0$ and $B$ scaled by 0.025. For the logic gates, we programmed in a 1-dimensional scalar dynamical system
\begin{equation}
\label{eq:logic}
\dot{x} = \frac{3}{13}x^3 - \left(123 + \frac{1}{13}\right)x + z,
\end{equation}
where $z$ is a function of $p$ and $q$ given by
\begin{equation*}
\begin{split}
z &= -0.1 + (p+0.1)(q+0.1)/0.2, \hspace{1cm} _{\mathrm{AND}}\\
z &=  0.1 + (p+0.1)(-q-0.1)/0.2, \hspace{1cm} _{\mathrm{NAND}}\\
z &=  0.1 + (p-0.1)(-q+0.1)/0.2, \hspace{1cm} _{\mathrm{OR}}\\
z &= -0.1 + (p-0.1)(q-0.1)/0.2, \hspace{1cm} _{\mathrm{NOR}},
\end{split}
\end{equation*}
and
\begin{equation*}
\begin{split}
z &=  -(p)(q)/0.1, \hspace{1cm} _{\mathrm{XOR}}\\
z &= (p)(q)/0.1, \hspace{1cm} _{\mathrm{XNOR}}.
\end{split}
\end{equation*}

\subsection{Parameters for ``Game Development on Recurrent Neural Architectures''}
The RC used to program our game of pong had $n = 400$ neurons, $\rho = 0$, and $B$ scaled by 0.1. The logic gates for the flip-flop were programmed using the same equations as in the previous section (Eq.~\ref{eq:logic}), multiplied by 20 to improve the speed of the response. 
\newpage

\section{Citation Diversity Statement}
Recent work in several fields of science has identified a bias in citation practices such that papers from women and other minority scholars are under-cited relative to the number of such papers in the field \cite{mitchell2013gendered,dion2018gendered,caplar2017quantitative, maliniak2013gender, Dworkin2020.01.03.894378}. Here we sought to proactively consider choosing references that reflect thediversity of the field in thought, form of contribution, gender, race, ethnicity, and other factors. First, we obtained the predicted gender of the first and last author of each reference by using databases that store the probability of a first name being carried by a woman \cite{Dworkin2020.01.03.894378,zhou_dale_2020_3672110}. By this measure (and excluding self-citations to the first and last authors of our current paper), our references contain 6.38\% woman(first)/woman(last), 10.64\% man/woman, 12.77\% woman/man, and 70.21\% man/man. This method is limited in that a) names, pronouns, and social media profiles used to construct the databases may not, in every case, be indicative of gender identity and b) it cannot account for intersex, non-binary, or transgender people. Second, we obtained predicted racial/ethnic category of the first and last author of each reference by databases that store the probability of a first and last name being carried by an author of color \cite{ambekar2009name, sood2018predicting}. By this measure (and excluding self-citations), our references contain 10.99\% author of color (first)/author of color(last), 15.25\% white author/author of color, 21.99\% author of color/white author, and 51.76\% white author/white author. This method is limited in that a) names and Florida Voter Data to make the predictions may not be indicative of racial/ethnic identity, and b) it cannot account for Indigenous and mixed-race authors, or those who may face differential biases due to the ambiguous racialization or ethnicization of their names.  We look forward to future work that could help us to better understand how to support equitable practices in science.